\documentclass[aps,twocolumn,pra,showpacs]{revtex4}
\usepackage[dvips]{graphicx}
\usepackage{dcolumn}
\usepackage{bm}
\def\be{\begin{equation}}
\def\ee{\end{equation}}
\def\bea{\begin{eqnarray}}          
\def\eea{\end{eqnarray}}
\def\bi{\begin{itemize}}
\def\ei{\end{itemize}}
\def\bin{\begin{enumerate}}
\def\ein{\end{enumerate}}

\def\la{\langle}
\def\ra{\rangle}
\begin{document}

\author{Bruno Eckhardt$^1$, Jakub S. Prauzner-Bechcicki$^2$, 
Krzysztof Sacha$^3$, and Jakub Zakrzewski$^3$}
\affiliation{$^1$Fachbereich Physik, Philipps-Universit\"at Marburg, 
D-35032 Marburg, Germany \\
$^2$ Faculty of Conservation and Restoration of Works of Art
Academy of Fine Arts, Smole\'nsk 9, Krak\'ow, Poland \\
 $^3$Marian Smoluchowski Institute of Physics and Mark Kac Complex Systems 
Research Centre, Jagiellonian University, Reymonta 4, 
30-059 Krak\'ow, Poland}

\title{Suppression of correlated electron escape in double ionization 
in strong laser fields}

\date{\today}

\begin{abstract}
The effect of the Pauli exclusion principle on double ionization of 
He atoms by strong, linearly polarized 
laser pulses is analyzed. We show that correlated electron escape, 
with electron momenta symmetric with respect to the field polarization axis, 
is suppressed if atoms are initially prepared in the  
metastable state $^3$S. The effect is a consequence of selection rules for 
the transition to the appropriate outgoing two-electron states.
We illustrate the suppression in numerical calculations of
electron and ion momentum distributions within a reduced dimensionality 
model.
\end{abstract}

\maketitle

The interaction of atoms with strong laser pulses of intensities in the range of 
$10^{14}$ W/cm$^2$ may result in the production of 
doubly charged ions. Early experiments indicated that the rates for this process
are much higher than expected on the basis of a single active electron model 
(see \cite{review} and references
therein). The commonly accepted explanation is based on the rescattering process,
where a temporarily ionized electron 
is driven back to the
atom by the electric field \cite{review,corkum93}. The acceleration by the field
provides additional energy which can then be transferred to the
second electron during the collision with the
residual ion. Measurements of the doubly charged ion and 
electron momentum distributions,
in addition to the total ionization yield, 
\cite{review,weber00n,weckenbrock0304} revealed that substantial number
of the electrons escapes with equal momenta along the field polarization axis.
The origin of this correlation was analysed as a consequence of 
the Coulomb repulsion between the outgoing electrons, their Coulomb 
attraction to the nucleus and the pulling force from the electric field
\cite{sacha01}. Except for the presence of the field during the 
decay of the complex formed in the rescattering
this analysis has many parallels to Wanniers \cite{wannier53,peterkop71rau71}
reasoning for 
double ionization after collision or photo excitation. Since in the
Wannier case it can be shown that the observed cross sections are constrained
by the symmetries of the problem \cite{klar76,greene82,review1,briggs95}, 
we here search for the corresponding effects in strong field double 
ionization.

Consequences of the exchange symmetry on double ionization yield 
have previously been studied \cite{guo00,ruiz04} within the aligned 
electron model \cite{aligned}. These authors found an reduction in
ionization rate in the antisymmetric state, but
overestimation of the Coulomb electron repulsion in
the aligned electron model
 did not allow them to treat the final
momentum distributions.
 These studies
were motivated by experimental observation of the difference  in 
single and double ionization of O$_2$ and N$_2$ molecules where the initial 
states are of triplet and singlet character, respectively \cite{guo98}.
Recent experiments \cite{eremina04,corkum05} observed differences 
between O$_2$ and N$_2$ molecules when
 the electron momentum correlations for components parallel to the 
light polarization axis were measured.

The constraints on the double ionization cross section arise from
 the symmetries of the wave functions for the outgoing electrons 
and can be derived from the study of the selection rules for 
transitions to two electron continuum states 
by Maulbetsch and Briggs \cite{briggs95}. 
They analyze transition matrix elements from an initial 
state $\psi$ to continuum final states of well defined electron momenta 
$\vec k_1$ and $\vec k_2$, i.e., $\la \vec k_1 \vec k_2|T|\psi\ra$ where 
$T$ is the appropriate transition operator.
As $r_1$, $r_2$ and $r_{12}=|\vec r_1-\vec
r_2|$ tend to infinity, the continuum states for a two electron
case of He approach the 
asymptotic behaviour 
$\la \vec r_1 \vec r_2|\vec k_1 \vec k_2\ra \rightarrow (2\pi)^{-3}\exp(i\vec
k_1\cdot\vec r_1+i\vec k_2\cdot\vec r_2+i\chi)$, where the phase 
$\chi=2\ln(k_1r_1+\vec k_1\cdot\vec r_1)/k_1+2\ln(k_2r_2+\vec k_2\cdot\vec
r_2)/k_2-\ln(kr_{12}+\vec k\cdot\vec r_{12})/(2k)$ with 
$\vec k=(\vec k_1-\vec k_2)/2$ accounts for the long range of Coulomb potentials.
These states and the full transition
matrix elements can now be projected onto a complete set of two electron 
states characterized by total angular momentum $L$, its projection $M$ on a 
chosen quantization axis $z$, total spin $S$ and any other necessary quantum 
numbers, collectively labelled $\alpha$,
\be
\la \vec k_1 \vec k_2|T|\psi\ra=\sum_{\alpha LMS}
\la \vec k_1 \vec k_2|\alpha LMS\ra
\la\alpha LMS|T|\psi\ra.
\label{matrix}
\ee 
Selection rules are now obtained for sets of  
$\vec k_1$, $\vec k_2$, $L$, $M$ and $S$ for which the overlaps 
$\la \vec k_1 \vec k_2|\alpha LMS\ra$ vanish identically. Various
situations are discussed in \cite{briggs95}. The case most appropriate
for correlated double escape with equal energy and equal momenta parallel
to the field axis is their case (I), where $k_1=k_2$,
both momenta lie in a plane, i.e., $|\phi_1-\phi_2|=\pi$, and their projections 
on the quantization axis coincide, $\theta_1=\theta_2$: then states with $(S+M)$ 
odd do not contribute to the transition matrix elements. Therefore, if
the initial wave function belongs to this subspace and if the evolution
operator preserves this subspace, these final states will be suppressed.

For the case of strong field double ionization 
the transition operator $T$ is the 
evolution operator corresponding to the Hamiltonian (in atomic units)
\be
H=\sum_{i=1}^2\left(\frac{\vec p_i^2}{2}-\frac{2}{r_i}\right)+
\frac{1}{|\vec r_1-\vec r_2|}+F(t)(z_1+z_2),
\label{fullham}
\ee 
where $F(t)$ contains the time dependence of a linearly polarized 
laser pulse. It is symmetric under exchange of electrons. Therefore, if
the initial wave function belongs to the $^3$S state of He, the sum in 
(\ref{matrix}) remains restricted to $M=0$ and $S=1$, since
both the $z$ component of the total angular momentum and the 
spin operators commute with the Hamiltonian. As a consequence,
the matrix element of the transition operator between the initial metastable
state and continuum states with $\vec k_1$ and $\vec k_2$ symmetric with respect
to the field polarization axis vanishes by the above selection rule. 
Thus, the differential cross section for
the double ionization process is exactly zero whenever the momenta 
of outgoing electrons are symmetric with respect to the field axis. 
Note that in experiments starting with the He ground state, i.e. $^1$S, 
and in the double ionization of noble gasses 
the symmetric escape is not forbidden --- 
correlated electron escapes are clearly visible in electron momentum 
distributions in the double ionization of Ne and Ar atoms
\cite{review,weber00n,weckenbrock0304}.

On the level of the classical dynamics of the two-electron system,
the selection rule eliminates a subset of phase space that includes
the $C_{2v}$ subspace of symmetric electron motion, which
has been shown to capture much of the observed momentum
distributions \cite{sacha01}.
In cylindrical coordinates, the $C_{2v}$ subspace is defined by
$R=\rho_1=\rho_2$, $p_R=2p_{\rho_1}=2p_{\rho_2}$, $Z=z_1=z_2$, 
$p_Z=2p_{z_1}=2p_{z_2}$, $|\phi_1-\phi_2|=\pi$ and $p_{\phi_1}=p_{\phi_2}=0$.
The Hamiltonian (\ref{fullham}) in the $C_{2v}$ subspace reduces to
\be
H=\frac{p_R^2+p_Z^2}{4}-\frac{4}{\sqrt{R^2+Z^2}}+\frac{1}{2R}+2ZF(t).
\label{c2v}  
\ee
The potential energy in (\ref{c2v}) [and obviously also the potential energy
in (\ref{fullham})] possesses a saddle located at 
$R=R_s\sin\theta_s$, $Z=R_s\cos\theta_s$, where 
$R_s=3^{1/4}/\sqrt{|F(t)|}$ and $\theta_s=\pi/6$
or $5\pi/6$ depending on the sign of $F(t)$. Electrons moving in the subspace
may simultaneously escape from the atom by going over the saddle. 
Trajectories living in the
full space that approach the saddle sufficiently symmetrically may lead to
simultaneous escape of electrons with  highly correlated final momenta
\cite{sacha01}.
For the static electric field (i.e. $F(t)=$const) the stability analysis of the
saddle allows one to estimate the energy dependence of the classical cross
section for correlated double escape for energy close to the saddle energy, 
i.e. to obtain
a counterpart of the Wannier threshold law in the presence of the electric
field \cite{eckhardt01}. Note that for the energy equal to the threshold value 
the only trajectory 
leading to the double escape corresponds to two electrons moving symmetrically  
in the same direction along the field axis. This is very different from
the original Wannier problem where the electrons escape symmetrically but in
opposite directions. 
The analysis of the quantum selection rule for double ionization of the 
metastable He atoms by linearly polarized laser pulses 
implies that, in terms of classical mechanics, 
this entire $C_{2v}$ symmetric subspace is forbidden for electrons.

With perfectly symmetric escape excluded, we have to look for escape with
almost symmetric electrons. Relaxing the condition on the angular degrees
of freedom, i.e. by allowing for arbitrary 
$\phi=\phi_1-\phi_2$ 
but $p_\phi=2p_{\phi_1}=-2p_{\phi_2}$ (i.e. $M=0$), we are led to consider
the invariant subspace of $C_v$ symmetry (the $C_v$ subspace contains
obviously the $C_{2v}$ subspace) \cite{sacha01}. 
Then the Hamiltonian of the system reads
\be
H=\frac{p_R^2+p_Z^2}{4}+\frac{p_\phi^2}{R^2}
-\frac{4}{\sqrt{R^2+Z^2}}+
\frac{1}{2R\left|\sin\left(\frac{\phi}{2}\right)\right|}+2ZF(t).
\label{cv}
\ee
Comparing (\ref{cv}) with (\ref{c2v}) we see that in the present case electrons
can perform a symmetric bending motion around the field axis.
In the $C_v$ subspace electrons may pass
over the saddle considered previously and ionize but their momenta need not  
lie in the plane $|\phi_1-\phi_2|=\pi$. In the quantum description there is
no selection rule for electrons with $k_1=k_2$ 
and with momenta that have the same projections on the polarization axis,
but with $|\phi_1-\phi_2|\ne\pi$. 

The consequences of the selection rule should show up most clearly 
in angle resolved double ionization cross sections, 
as they result in a zero for 
the symmetric
momenta and an azimuthal angle difference of $\pi$. However, it will 
also show up as a reduction in intensity 
in the angle integrated cross sections, like the ones where
only the parallel momenta components of ionizing electrons are measured
\cite{review,weber00n}. 
We will demonstrate this with calculations in
1+1-dimensional model below, but we emphasize that
the cleanest verification is possible with fully momentum 
resolved cross sections as obtained in recent experiments 
\cite{weckenbrock0304}.

Full three-dimensional (3D) calculations of the double ionization process 
are barely feasible
\cite{taylor,faisal}, so we take recourse to low-dimensional models. 
Aligned electron models \cite{aligned}
are not suitable for our purpose since there the
symmetric simultaneous escape of electrons is forbidden by 
the overestimated electron repulsion. The 1+1-dimensional (1+1D) model 
proposed in \cite{eckhardt06} does allow for symmetric escape
by confining the electrons to move along one-dimensional (1D) tracks that form 
angles $\pi/6$ and $5\pi/6$ with respect to
the field polarization axis.  
The Hamiltonian of the system, in atomic units, reads
\bea
H_{1D}&=&\sum_{i=1}^2\left(\frac{p_{R_i}^2}{2}-\frac{2}{|R_i|}\right)
+\frac{1}{\sqrt{(R_1-R_2)^2+R_1R_2}} \cr
&&+\frac{F(t)\sqrt{3}}{2}(R_1+R_2),
\label{ham1d}
\eea
where $R_1$ and $R_2$ denote positions of electrons on the chosen tracks and 
$p_{R_1}$ and $p_{R_2}$ are conjugate momenta.
In this model the rescattering process, single and sequential double ionization 
are present but importantly also the symmetric simultaneous escape of electrons 
can be properly simulated \cite{jqb07}. Indeed, in the 3D case, there is a 
single saddle around which two electrons can escape simultaneously and 
that saddle possesses two unstable directions. The 1+1D model also has a 
saddle located at the same place as in the 3D case, $R_1=R_2=-{\rm sgn}[F(t)]R_s$, 
also with two unstable directions. Hence, the key topology elements 
of the phase space of the 3D case are properly reproduced in the model. 
The symmetric subspace of the classical phase space 
in the model is defined by $R_1=R_2$ and $p_{R_1}=p_{R_2}$. 
If, in quantum simulations, we start with the ground state of the unperturbed 
system symmetric electron escape is not forbidden. However, if we
chose as an initial state the first excited state, which is antisymmetric 
with respect to electron exchange (in analogy to the $^3$S state)
the wavefunction vanishes at $R_1=R_2$. The symmetric subspace is then
forbidden.

We have performed numerical integration of the Schr\"odinger equation
corresponding to the Hamiltonian (\ref{ham1d}) with Coulomb singularities 
smoothed by the replacement $1/x\rightarrow 1/\sqrt{x^2+e}$ with $e=0.6$, 
starting with the ground and first excited states of the unperturbed 
Hamiltonian. The ground state energy, for the chosen parameter $e$ 
is $-2.83$ while the first excited state is at $-2.21$.
We have used a single cycle pulse of the form 
$F(t)=F_0f(t)\sin(\omega t)$, where the pulse envelope 
$f(t)=\sin^2(\omega t/2)$, with $\omega=0.06$ and $F_0=0.15$.
Single ionization in the case of the initial first excited state 
is very efficient (with probability equal to 0.98) and much 
more probable than in the case of the ground state 
(with probability 10$^{-4}$).  
That is due to the higher energy of the first excited state, 
which allows electrons to easily pass over the Stark barrier 
and to escape near the peak intensity of the field.
Probabilities for double ionization are comparable in both
cases, about 10$^{-6}$,
thus indicating that the difference in initial energy does not have an
exessively strong effect on double ionization. 

\begin{figure}
\includegraphics[width=0.38\textwidth]{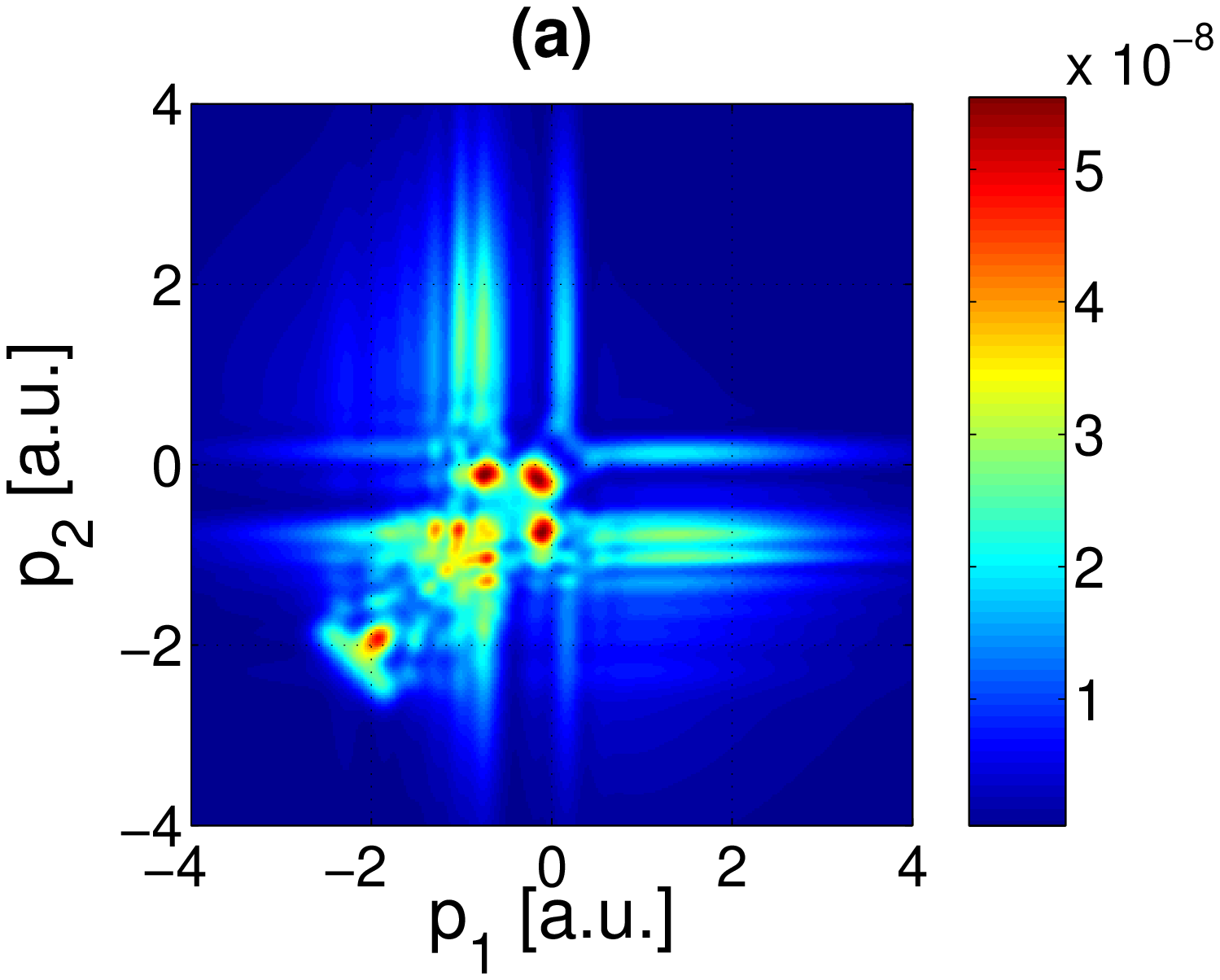}
\includegraphics[width=0.38\textwidth]{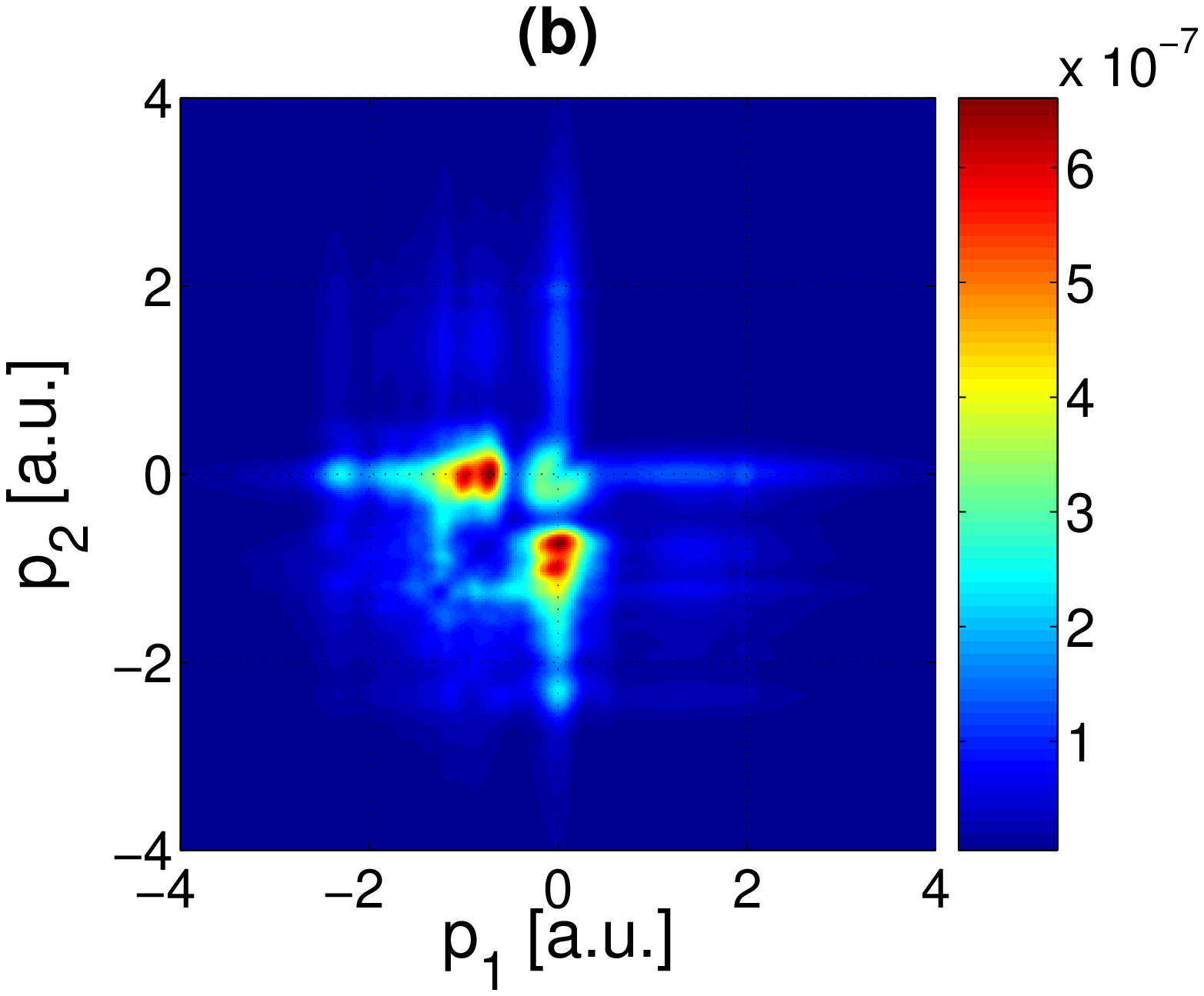}
\caption[]{
Electron momentum distributions for double ionization in strong linearly
polarized laser field starting from the ground state of the unperturbed atom
(a) and from first excited state (b). 
The distributions are obtained from Fourier transforms of the parts of the 
final wavefunctions in the regions 
$|R_1|$, $|R_2|>50$ and are averaged with 
a Gaussian of width 0.07 in order to model experimental resolution. 
The time-dependance of the field is $F(t)=F_0f(t)\sin(\omega t)$ where
$F_0=0.15$, $\omega=0.06$ and one-cycle pulse duration with an envelope
$f(t)=\sin^2(\omega t/2)$.
The concentration of the outgoing
momenta in one quadrant is a result of the single cycle laser pulse.
The low intensity along the diagonal in panel (b) is the signature
of the suppression of symmetric double escape by the selection.
\vspace{-0.5cm}
\label{one}}
\end{figure}

\begin{figure}
\includegraphics[width=0.41\textwidth]{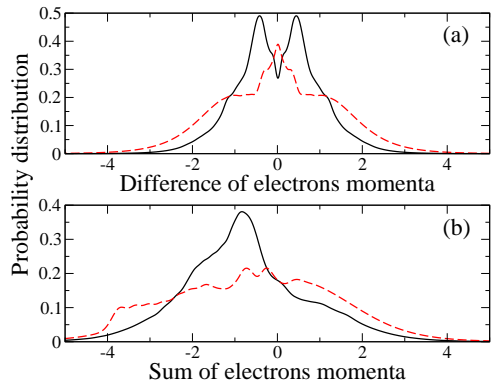}
\caption[]{
The probability distributions for differences, $p_{R_1}-p_{R_2}$,  [panel (a)], 
and for sums, 
$p_{R_1}+p_{R_2}$,  [panel (b)], of electron momenta for double 
ionization in a strong linearly polarized laser field.
Data for initial ground state (red dashed curves) are compared with those
obtained for the initial first excited state (solid black lines).
The minimum in the difference distribution in panel (a) reflects 
the suppresion of symmetric escapes.
All parameters are the same as in Fig.~\ref{one}.
\vspace{-0.5cm}
\label{two}}
\end{figure}

Figure~\ref{one} shows electron momentum distributions in the 
double ionization of He for the initial ground and first excited 
states. The distributions have been obtained adopting the method 
proposed in \cite{engel} as described in \cite{jqb07}. 
The momentum distributions are obtained by Fourier transforms 
of parts of the final wavefunctions in the regions 
$|R_1|$, $|R_2|>50$ and they are Gaussian (of width 0.07) averaged in 
order to mimic the experimental resolution. 
For double ionization starting from the symmetric ground state
there is a symmetric distribution of parallel final momenta with
a maximum along equal momenta, $p_{R_1}=p_{R_2}$. That the distribution
reaches only into the negative momentum sector is due to the short
pulse which contains only one cycle, \cite{liu04}.
When the initial state is the antisymmetric
state, the corresponding distribution shown in 
Fig.~\ref{one}b has a clear minimum along the axis $p_{R_1}=p_{R_2}$,
reflecting the absence of the perfectly correlated escape,
as predicted by the selection rule.
The small residual contribution at the center is due
to averaging with the Gaussian that models
experimental resolution. 

The signatures of the effect can be enhanced, as shown
in Fig.~\ref{two}a, by considering the
distributions of the momenta differences, $p_{R_1}-p_{R_2}$.
One finds a strong maximum at $p_{R_1}-p_{R_2}=0$
for double ionization starting from the symmetric ground state 
and a profound minium when the initial state is the antisymmetric
first excited state. Note again that the minimum in 
Fig.~\ref{two}a does not reach zero
because of the averaging. But  
Fig.~\ref{two}a clearly demonstrates that the resolution 
attainable in current experiments 
is sufficient to observe effects of the selection rule we analyze here. 

In Fig.~\ref{two}b the distributions of momenta sum (equivalent to ion 
momentum distributions), i.e. $p_{R_1}+p_{R_2}$, 
in double ionization of He atoms are plotted. We see that the  
momenta sum distribution shows no simple signature of the absence of 
highly correlated electron escape in the case of the initial first excited
state except, possibly a small narrowing of the distribution. 
By comparison, the existence of correlated pairs of electrons for 
the initial ground state case broadens the
corresponding sum distribution (dashed line). 
We also would like to point out that the asymmetry 
of the ion momentum distribution seen in Fig.~\ref{two}b 
reflects the carrier-envelope phase as expected \cite{liu04}.

Finally, we would like to stress that the numerical results presented 
are obtained in the reduced dimensionality model where the 1D electron tracks 
are chosen in two-dimensional space containing the polarization axis. Direct
application of the reduced 1+1D model to experimental data is thus
possible only if events with electron momenta lying in a plane that contains the
field polarization axis are collected only.

We have shown that the selection rules of Maulbetsch and Briggs 
\cite{briggs95} imply a difference in the final momenta distributions
in double ionizatoin by strong linearly polarized laser pulses
depending on the symmetries of the initial state. 
For double ionization in He starting from the 
metastable state $^3$S correlated electron escape 
with final electron momenta symmetric with respect to 
the polarization axis is suppressed. Numerical simulations in
a reduced dimensionality model show a clear dip in the momentum
distributions, and demonstrate that the effect appears despite
differences in the initial energy. The effect should be within
reach of todays experiments with their resolution
and detailed kinematic characterization of the final state.
Extensions to double ionization in molecules are immediate. A prime candidate
are O$_2$ molecules with triplet initial state oriented along the polarization
axis. Recent experiments with aligned N$_2$ molecules 
\cite{corkum05} probed a singlet initial state only.

The work was supported by Polish Government scientific funds 
(2005-2008) as a research project, by Marie Curie ToK project 
COCOS (MTKD-CT-2004-517186), and Deutsche Forschungsgemeinschaft.

\end{document}